\documentstyle[twoside,fleqn,espcrc2,epsf]{article}
\newcommand{\AmS}{{\protect\the\textfont2
  A\kern-.1667em\lower.5ex\hbox{M}\kern-.125emS}}

\title{
{\vspace{-2.5cm} \normalsize
\hfill \parbox{40mm}{CERN/TH-98-287}\\
\hfill \parbox{40mm}{MPI-PhT/98-67}}\\[12mm]
How the PHMC algorithm samples configuration space
\thanks{talk presented by R.~Frezzotti at the
XVI International Symposium on Lattice Field Theory, 
Boulder, Colorado, USA}
}

\author{
R.~Frezzotti\address{Max-Planck-Institut f\"ur Physik, \\
F\"ohringer Ring 6, D-80805 M\"unchen, Germany}
and K.~Jansen\address{CERN, \\
1211 Gen\`eve, Switzerland}
}

\begin{document}

\begin{abstract}
We show that in practical simulations of lattice QCD with
two dynamical light fermion species the PHMC algorithm samples 
configuration space differently from the commonly used
HMC algorithm.
\end{abstract}

\maketitle

%////////////////////////////////////////////////////////////////////////////
\setcounter{footnote}{0}

%////////////////////////////////////////////////////////////////////////////
\section{Introduction}

Simulations of lattice QCD with light dynamical fermions still pose
severe problems, concerning both the computational cost and ergodicity 
properties. The PHMC algorithm \cite{phmc_intro}
represents an attempt of improving
on both points over the commonly used HMC algorithm \cite{hmc}.

The main idea \cite{defo_taka} 
of the PHMC algorithm is to sample gauge configuration
space -through a standard HMC method- by using an approximate action
where, in the fermion sector, the inverse of the squared Hermitian
Dirac operator ($Q^2 \propto M^\dagger M$) is replaced by a suitable 
polynomial approximant, $P_{n,\epsilon}(Q^2)$.
The use of a well controlled polynomial approximation makes also
possible to correct for it through 
an efficient reweighing technique \cite{phmc_intro},
leading to exact (reweighted) sample averages for lattice QCD with 
$n_f =2$ degenerate fermion species:
\begin{equation} \label{QCD_ave}
\langle {\cal O} \rangle = \langle W \rangle_{PHMC}^{-1}
\langle {\cal O} W \rangle_{PHMC} \;
\end{equation}
where ${\cal O}$ stands for any observable and $W$ is a noisy
estimate of $\det [Q^2 P_{n,\epsilon}(Q^2)]$.

According to the Chebyshev approximation method, the polynomial
in $s$ having degree $n$, $P_{n,\epsilon}(s)$, approximates the
function $1/s$, with $s>0$, with a relative error that is bounded
by $\delta \simeq 2 \exp (-2\sqrt{\epsilon}n)$ in the range\footnote
{The operator $Q^2$ is assumed to be normalized so that its 
highest eigenvalues is always smaller than $1$.}
$\epsilon \le s \le 1$ and that quickly increases as $s$ gets smaller 
than $\epsilon$. As a consequence,
in the molecular dynamics (MD) update, the role of the lowest
eigenvalues of $Q^2$, denoted here by $\lambda_{\rm min}$,
is taken by $\epsilon$, which can be chosen -in practice- about 
$2 \langle \lambda_{\rm min} \rangle$. This leads to a computational cost
per MD trajectory correspondingly smaller -by about a factor 2- than
in the case of the HMC algorithm. 
%Moreover, the approximate action
%used in the MD evolution gives to gauge configurations carrying very
%low lying (compared to $\epsilon$) modes of $Q^2$ a much larger probability
%to be generated than in the HMC algorithm. 
Moreover, the approximate action used in the MD evolution makes gauge
configurations with very low (compared to $\epsilon$) eigenvalues of
$Q^2$ to be generated with much higher probability than in the HMC algorithm.
These gauge configurations
are expected to be important for the sample average of many fermion
observables and for associated changes of topological sectors.
We then expect the PHMC algorithm to show -in critical
situations- ergodicity properties that are different from the 
ones of the HMC algorithm.

%////////////////////////////////////////////////////////////////////////////
%\section{Test studies and sampling of configuration space}
\section{Main results of PHMC tests}
\label{sec:results}

The PHMC algorithm has been implemented on APE computers and carefully
studied \cite{phmc_I}\cite{phmc_II}; particular care has been devoted
to the role of reweighing and to the tuning of $n$ and $\epsilon$, 
for which a practical procedure was suggested \cite{phmc_II}.
All test studies were performed on lattices with Schr\"odinger functional
(SF) boundary conditions \cite{sf_biblio}, which enabled us to work at
vanishing quark mass. We have used Wilson fermions
-in both the standard and the O(a) improved versions- with {\em even--odd}
preconditioning and a Sexton--Weingarten integration scheme for the
MD evolution. 
Within the statistical uncertainties, we have always found consistent results
for the mean values of several pure gauge and fermion observables 
obtained from the HMC and the PHMC algorithm.
%
%The interplay between polynomial approximation of $(Q^2)^{-1}$ and noise
%induced by the reweighing procedure has been investigated, leading to
%an empirical criterion for the choice of $n$ and $\epsilon$; practically
%satisfactory ways to deal with memory requirements and rounding-error
%problems in critical situations are also available \cite{phmc_I}. 

%% PHMC : performance tests summary
%\begin{table*}[bt]
\begin{table}[bt]
\setlength{\tabcolsep}{0.3pc}
\newlength{\digitwidth} \settowidth{\digitwidth}{\rm 0}
\catcode`?=\active \def?{\kern\digitwidth}
\caption{
The average condition number, $\langle k \rangle$, of the preconditioned 
fermion matrix and the costs, $C_{Q\phi}$, of a MD trajectory 
in performance tests at $\kappa = \kappa_c$. 
More details in refs.\hspace{1mm}[1] (test $a$) and [5] (tests $b$,$c$).
%We omit here many tests on $4^4$ lattices [4].
%$\kappa$ was set to 0.1585,  0.1343, 0.1379 for tests a, b, c respectively.
}
\label{tab:perfo_tests}
\begin{tabular}{lllllll}
\hline
\phantom{a} & Lattice & $\beta$ & $c_{\rm sw}$ &
$\langle k \rangle$   & $C_{Q\phi}^{\rm HMC}$  &  $C_{Q\phi}^{\rm PHMC}$ \\
\hline \hline
$a$:   &  $8^4$  & $5.6$ & $0$  & 
$ 720$ &  7398   &  3974   \\
$b$:   &  $8^3 \cdot 16$  & $6.8$  & $1.4251$  & 
$ 760$ &  7750   &  5956   \\
$c$:   &  $8^3 \cdot 16$  & $5.4$  & $1.7275$  & 
$1500$ & 19734  &  11450  \\
\hline \hline
\end{tabular}
\vspace{-0.3cm}
%\end{table*}
\end{table}

Performance tests\footnote{We omit to discuss here
many tests on $4^4$ lattices \cite{phmc_I}.} against the HMC algorithm 
have been performed at {\em vanishing} quark
mass on lattices of small and intermediate physical size, with spatial
length never larger than $1$ fm. This physical situation -although very
different from the ones where most unquenched simulations are performed-
is of practical interest for non perturbative renormalization studies.
We summarize in table \ref{tab:perfo_tests} the main results of our
performance tests:
the computational costs $C_{Q\phi}^{\rm HMC}$ and 
$C_{Q\phi}^{\rm PHMC}$ are given in units of 
fermion matrix ($Q$) times pseudo-fermion vector ($\phi$) multiplications
and refer to a full trajectory, accounting in the PHMC case
also for the reweighing procedure. However only for tests $a$ and $b$
comparing these costs corresponds to a comparison of the actual costs to
generate an independent gauge configuration, 
since for almost all the considered
observables compatible errors (within O($15\% $) relative uncertainties) 
are obtained from the two algorithms when
the same statistics is employed. In the case of test $c$ the accumulated 
statistics is not enough to make
any definite statement on statistical errors and corresponding uncertainties.
Moreover a direct performance comparison is made problematic by the
significant differences in sampling the configuration space that we
are going to discuss.

\begin{figure}[htb]
\vspace{-10mm}
\hspace{-5mm}
\begin{center}
\leavevmode
\epsfxsize=8.0cm
\epsfysize=6.0cm
\epsfbox{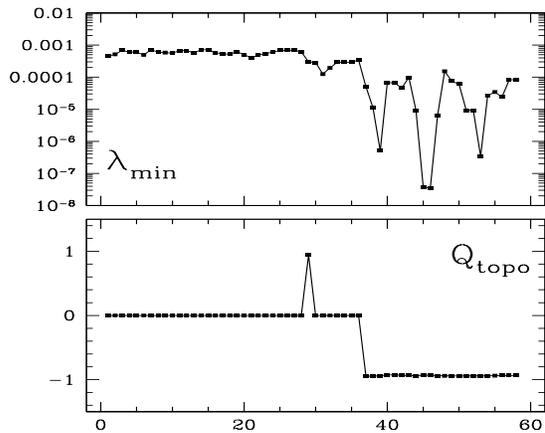}
\end{center}
\vspace{-15mm}
\caption{
An example of Monte Carlo time evolution for $\lambda_{\rm min}$ and
the naive topological charge after cooling, $Q_{\rm topo}$. 
}
\label{fig:Q_histo}
\vspace{-20pt}
\end{figure}

As shown in fig.~3 of ref.~\cite{phmc_I} and fig.~4 of ref.~\cite{phmc_II},
the PHMC sample in the case of test $c$ includes a significant fraction of
configurations carrying one (or few) isolated modes of the fermion matrix
lying some orders of magnitude below the average value of $\lambda_{\rm min}$.
Moreover, different values of the naive topological 
charge are measured after 500 cooling iterations, suggesting that changes
of topological sectors do occur - even at zero quark mass and in a space
volume less than $1$ fm$^3$. A typical example is shown in 
fig.~\ref{fig:Q_histo}. We recall that no index theorem has
to hold in this case and refer to \cite{phmc_II} for further considerations. 
No exceptionally small eigenvalues 
were observed in the corresponding HMC simulation \cite{jansom}. 

\begin{figure*}[htb]
\vspace{-5mm}
\begin{center}
\leavevmode
\epsfxsize=16.0cm
\epsfysize=10.0cm
%%%\epsfbox{fa_replicum3qh1.ps}
%%%\epsfbox{fa2_replicum3qh1.ps}
\epsfbox{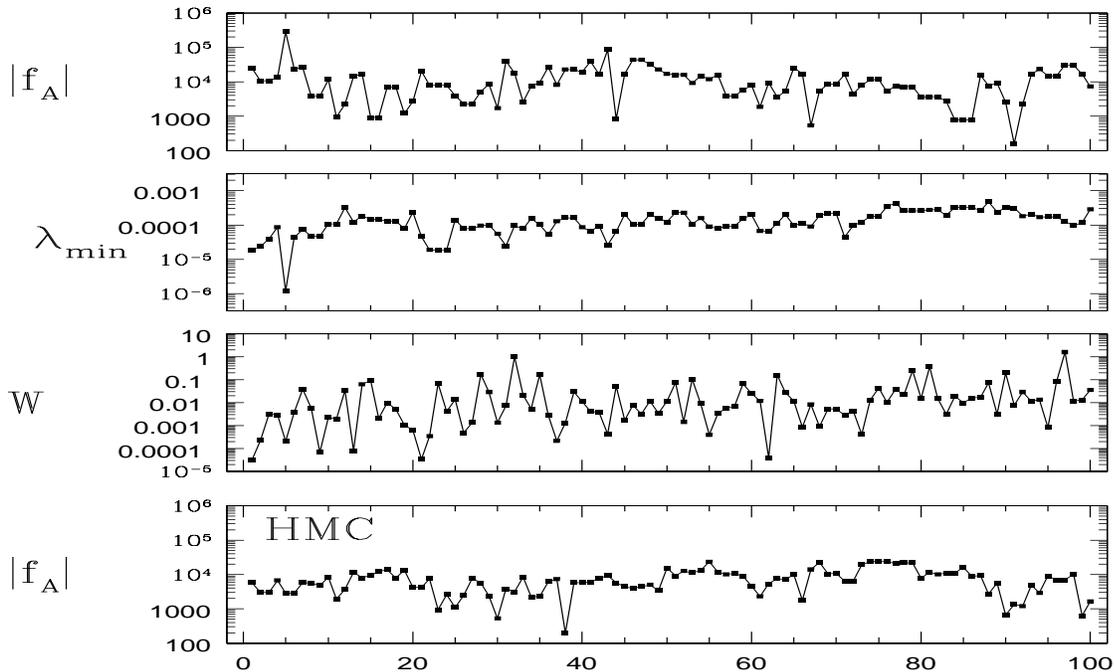}
\end{center}
\vspace{-20mm}
\caption{
A  segment of the Monte Carlo history of $f_A(T/2)$ as obtained from the
PHMC and the HMC algorithms. In the PHMC case we also show the corresponding
history of $\lambda_{\rm min}(\hat{Q}^2)$ and $W$. 
}
\label{fig:fa_histo}  
\vspace{-0pt}
\end{figure*}

We present in fig.~\ref{fig:fa_histo} the Monte Carlo time evolution
of the four--fermions Green function $f_A(T/2)$ (see e.g. eq.(14) of
ref.~\cite{phmc_II}
for its definition), taking two typical Monte Carlo history
segments from our PHMC
and HMC data. In the PHMC case we see
that peaks for $f_A(T/2)$
occur in coincidence with very small values of $\lambda_{\rm min}$. The 
contributions to QCD sample averages, eq.(\ref{QCD_ave}),
coming from these ``exceptional'' configurations,
are made of ``normal'' size by the corresponding small 
values of $W$. Many examples of this
behaviour, with even larger spikes in the values of $f_A(T/2)$
and $\lambda_{\rm min}$, were seen in the PHMC data.
A similar behaviour has been also observed
in numerical studies of Supersymmetry \cite{susy_mont}.
On the other hand, the HMC algorithm seems 
to generate with very low probability these exceptional
configurations, from which a relevant finite contribution to QCD
sample averages for many fermion observables may in principle come.

\section{Conclusions}

We have shown that the PHMC algorithm, as expected, samples the gauge 
configuration space differently from the HMC algorithm, while being at
least competitive with it from the performance point of view. If gauge
configurations carrying exceptionally small eigenvalues of the fermion
matrix are important for some observables, the PHMC algorithm should be
largely superior, owing to its ability\footnote{
This remains true even in presence of exact fermion
zero modes, as discussed in \cite{phmc_I}.} of 
sampling and properly treating these configurations. 
%We think that ergodicity and performance properties of the HMC and
%the PHMC algorithm deserve to be further studied, with more statistics and 
%on larger volumes.
We think that consequences of the different ergodicity and performance
properties of the HMC and the PHMC algorithm deserve to be further studied
with more statistics and on larger volumes.

This work is part of the ALPHA collaboration research programme. We thank
DESY for allocating computer time to this project.

%////////////////////////////////////////////////////////////////////////////

%
\end{document}